\documentclass[journal,twoside,web]{ieeecolor}
\usepackage{jsen}
\usepackage{cite}
\usepackage{amsmath,amssymb,amsfonts}
\usepackage{algorithmic}
\usepackage{graphicx}
\usepackage{textcomp}
\usepackage{wrapfig}
\usepackage{epstopdf}
\usepackage{array}
\usepackage{subfigure}
\usepackage[subfigure]{tocloft}
\usepackage{bm}
\usepackage{setspace}
\usepackage{easyReview}
\usepackage{amsfonts}
\usepackage{lipsum,amsmath}
\usepackage{amssymb}
\usepackage{txfonts}
\usepackage{float}
\usepackage{graphicx}
\usepackage{epstopdf}
\usepackage{enumerate}
\usepackage{algorithm}
\usepackage{algorithmic}
\usepackage{amsmath} 
\usepackage{booktabs} 
\usepackage{bbding}
\usepackage{gensymb}
\usepackage{amsmath}
\UseRawInputEncoding

\usepackage{cuted}\stripsep -2pt plus 2pt minus 1pt

\def\BibTeX{{\rm B\kern-.05em{\sc i\kern-.025em b}\kern-.08em
    T\kern-.1667em\lower.7ex\hbox{E}\kern-.125emX}}
\definecolor{abstractbg}{rgb}{0.89804,0.94510,0.83137}
\begin{document}
	\begin{spacing}{1.0}
\title{Integrated Sensing and Communication Neighbor Discovery for MANET with Gossip Mechanism}
\author{Zhiqing Wei,
	    Chenfei Li,
	    Yanpeng Cui,
	    Xu Chen,
	    Zeyang Meng,
	    and Zhiyong Feng
\thanks{This work was supported in part by the National Natural Science Foundation of China (NSFC) under Grant 92267202, in part by the National Key Research and Development Program of China under Grant 2020YFA0711302, and  in part by the National Natural Science Foundation of China (NSFC) under Grant U21B2014, and Grant 62271081.}
\thanks{Zhiqing Wei, Chenfei Li, Yanpeng Cui, Xu Chen and Zeyang Meng are with Key Laboratory of Universal Wireless Communications, Ministry of Education, School of Information and Communication Engineering, Beijing University of Posts and Telecommunications (BUPT), Beijing 100876, China (e-mail: weizhiqing@bupt.edu.cn, chenfeili@bupt.edu.cn, cuiyanpeng94@bupt.edu.cn, chenxu96330@bupt.edu.cn, mengzeyang@\\bupt.edu.cn, fengzy@bupt.edu.cn).}
\thanks{Corresponding authors: Zhiyong Feng; Chenfei Li.}}
\IEEEtitleabstractindextext{%
\fcolorbox{abstractbg}{abstractbg}{%
\begin{minipage}{\textwidth}%
\begin{wrapfigure}[12]{r}{2.1in}%
\includegraphics[scale=0.325]{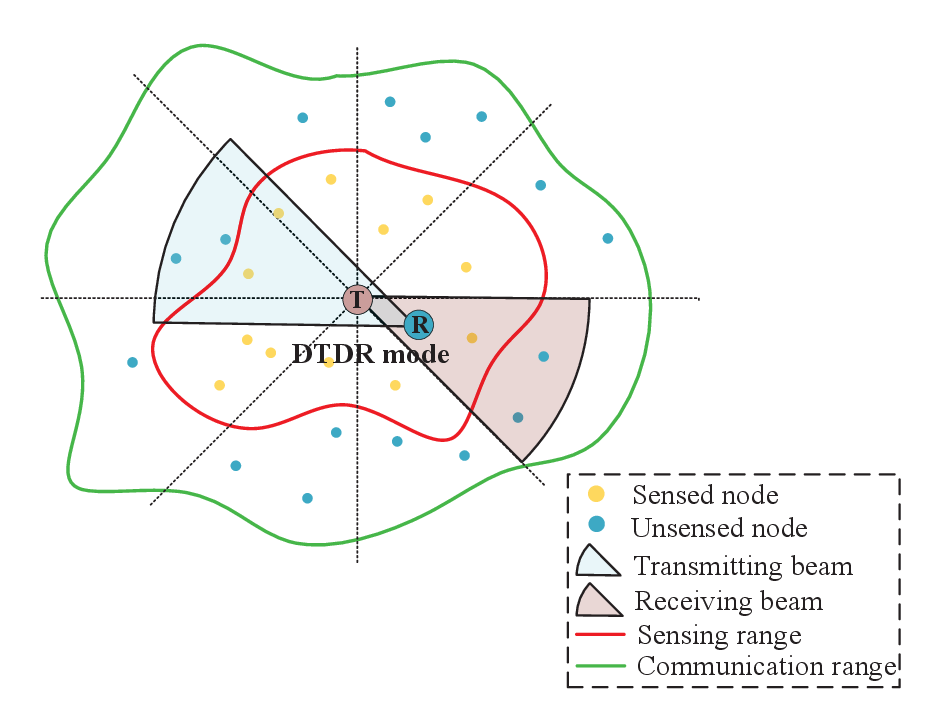}%
\end{wrapfigure}%
\begin{abstract}
Mobile Ad hoc Network (MANET), supporting Machine-Type Communication (MTC), has a strong demand for rapid networking. Neighbor Discovery (ND) is a key initial step in configuring MANETs and faces a serious challenge in decreasing convergence time. Integrated Sensing and Communication (ISAC), as one of the potential key technologies in the 6th Generation (6G) mobile networks, can obtain the sensing data as the priori information to accelerate ND convergence. In order to further reduce the convergence time of ND, this paper introduces the ISAC-enabled gossip mechanism into the ND algorithm. The prior information acquired by ISAC reduces the information redundancy brought by the gossip mechanism and thus decreases the probability of collision, which further improves convergence speed. The average number of discovered nodes within a given period is derived, which is applied as the critical metric to evaluate the performance of ND algorithms. The simulation results confirm the correctness of the theoretical derivation and show that the interplay between the prior mechanisms and the gossip mechanism significantly reduces the convergence time. In addition, to solve the problem of imperfect sensing information, reinforcement learning is applied. Under the constraints of the convergence condition, the non-Reply and non-Stop Algorithm based on Gossip and Q-learning (GQ-nRnS) proposed in this paper not only ensures the completeness of ND, but also maintains a high convergence speed of ND. Compared with the Q-learning-based ND algorithm (Q-ND), the average convergence time of the GQ-nRnS algorithm is reduced by about 66.4\%.
\end{abstract}

\end{minipage}}}

\maketitle

\section{Introduction}
\label{section_1}
\subsection{Background}

\IEEEPARstart{R}{ecently}, Wireless Ad hoc Network (WANET), as a common network supporting the Internet of Things (IoT), has been widely applied in the fields such as disaster relief, emergency rescue, exploration and monitoring, which has attracted wide attention both in academia and industry \cite{1,2,3}. Due to the deployment flexibility and self-healing potential \cite{6,7}, Mobile Ad hoc Network (MANET), as a class of WANET, supports a large number of applications in scenarios of Machine-Type Communication (MTC), such as Vehicular Ad hoc Network (VANET) and Flying Ad hoc Network (FANET). The frequent topology changes in MANET have led to the demand for fast networking. Fast and accurate Neighbor Discovery (ND) works as a critical process for mobile networking, and can accelerate the implementation efficiency of upper-layer protocols such as medium access and routing protocols, meeting the demand for fast networking \cite{8}.

Compared with omnidirectional antennas, directional antennas have extra advantages, including high transmission capacity \cite{9}, long transmission range \cite{10}, strong anti-interference ability \cite{11}, etc., making them suitable for MANET. However, the application of directional antennas also brings challenges for beam alignment, which significantly deteriorates the convergence time of ND algorithms.

With directional antennas, ND algorithms can be classified into two categories: Completely Random Algorithms (CRA) and Scan-Based Algorithms (SBA) \cite{12}. CRA randomly selects the transceiver state and the beam direction with a certain probability at the beginning of each time slot. SBA selects the beam direction in turn according to a predefined sequence in each time slot, and then selects the transceiver state at random or as determined by the probability model. SBA requires that the relative locations of nodes within one scan cycle should be the same \cite{12}, while CRA only requires the relative locations of nodes to remain unchanged within a time slot \cite{13}, which is more applicable for MANET. However, the CRA has the long-tail convergence problem due to beam misalignment, where two neighbor nodes may not discover each other for a long time, directly extending convergence time of ND \cite{14}.

To reduce the convergence time of ND, existing literature has conducted research on providing prior information, designing interaction mechanisms and introducing machine learning algorithms, which are reviewed as follows.

\subsubsection{Prior information} The prior information allows nodes to know neighbor information in the network in advance, which avoids exploring neighbors blindly in all directions and reduces the convergence time of ND. The prior information could be available through internal system design and external sensor sensing. {\em In terms of internal system design}, Burghal {\em et al.} in \cite{15} exploited the prior neighbor information that obtained at lower-frequency band to assist ND process at the upper-frequency band. Sorribes {\em et al.} in \cite{16} employed the collision detection mechanism to gather information for adapting node states, thereby reducing the convergence time of ND. {\em In terms of external sensor sensing}, Wei {\em et al.} in \cite{43} utilized vehicle distribution sensed by roadside units as the priori information to assist ND. Furthermore, Integrated Sensing and Communication (ISAC) technology integrates radar sensing into the communication module, improving the utilization of spectrum and hardware resources \cite{21,22}. Borrowing ISAC, nodes can accurately sense the distribution of surrounding neighbor nodes in real time, and they can make pre-determined decisions regarding packet transmission and the direction of the beam for transmitting those packets, which reduces the network resource overhead and ND convergence time compared with the blind search process of traditional ND algorithms. \cite{24} and \cite{25} introduced the prior information provided by radar sensing to improve the SBA algorithm. Liu {\em et al.} in \cite{26} determined the number and location information of neighbors with the double-face phased array radar, which is used as the priori information to accelerate convergence of ND. Wei {\em et al.} in \cite{27} designed four radar-assisted ND algorithms based on the accuracy of the prior information. 

\subsubsection{Interaction mechanism} The goal of ND is to obtain neighbor information directly or indirectly via the packet exchange, establishing the connection between neighbors. The interaction mechanism is designed to allow nodes to exchange packets as rapidly as possible in order to achieve fast convergence of ND. The gossip mechanism plays an important role. The gossip mechanism can randomly select relay nodes to forward data, which reduces the network overhead (the number of control messages) and improves the reliability of the network to some extent \cite{38}. When applied in ND, the nodes can indirectly learn about neighbor information by interacting with relay nodes, which substantially reduces the convergence time of ND. \cite{39} and \cite{40} demonstrated that the gossip-based ND algorithm obviously converges faster than the direct ND algorithms. Astudillo {\em et al.} in \cite{41} obtained location information of neighboring nodes through gossip packets to speed up the convergence of ND. Karowski {\em et al.} in \cite{42} compared the performance of gossip-based ND with that of multiple-transceiver-based ND and proved that the convergence performance of the gossip-based ND is equivalent to that of ND using 2 to 4 transceivers simultaneously.

\subsubsection{Machine learning algorithms} Recently, machine learning algorithms such as model prediction and reinforcement learning are widely applied in ND. {\em In terms of model prediction,} the prediction model utilizes the observed information to estimate or predict the location of the target to assist ND. Li {\em et al.} in \cite{28} applied the continuously updated autoregressive mobility model to predict locations of the moving nodes to facilitate ND. Liu {\em et al.} in \cite{29} combined Kalman filter prediction and real location broadcasting to achieve fast and accurate ND. {\em In terms of reinforcement learning,} reinforcement learning optimizes ND through the interaction between the agent and the environment. On value-based models, \cite{33} and \cite{34} introduced the Q-learning model into ND and designed suitable reward and punishment policies to accelerate convergence of ND. On policy-based models, \cite{36} and \cite{37} modeled the ND as a learning automaton and the nodes will adjust the direction of the directional antennas according to the learning experience, achieving fast convergence. 

Overall, although the convergence time of ND is reduced, there are still challenges for ND algorithms. 1) Information redundancy: The fast convergence of the gossip-based ND algorithms is obtained at the cost of increasing information redundancy, which causes network overload and congestion. 2) Imperfect sensing information: The perfect prior information was adopt to reduce the convergence time of ND in the existing literature. However, the imperfect prior information will affect the convergence of ND.

\subsection{Our Contributions}
In this paper, the interplay between the priori information and the gossip mechanism avoids information redundancy and further facilitates fast convergence of the ND algorithm. In addition, reinforcement learning is also introduced to address the problem of imperfect prior information. The main contributions of this paper are summarized as follows.

\begin{itemize}
	\item[1.] ISAC enabled gossip mechanism: This paper proposes four ISAC ND algorithms with gossip. The prior information obtained by ISAC is applied to the gossip mechanism, reducing the redundant information in the traditional gossip mechanism and solving the long-tail latency problem. The average number of discovered nodes within a given period is derived, which is adopted as the critical metric to evaluate the performance of the above four algorithms.
	
	\item[2.] Reinforcement learning mechanism: The Q-learning mechanism is further applied in the designed ISAC-enabled ND algorithm, which adaptively selects the beam direction by the reward and punishment mechanism, solving the problems of convergence delay caused by imperfect sensing information.
\end{itemize}

\subsection{Outline of This Paper}
The rest of this paper is organized as follows. Section \ref{section_2} describes the system model. Section \ref{section_3} proposes four ISAC ND algorithms with gossip and analyzes the performance of the four algorithms. Section \ref{section_4} considers imperfect sensing information and proposes an ISAC-enabled ND algorithm with gossip and reinforcement learning. Section \ref{section_5} simulates and analyzes the proposed algorithms. Section \ref{section_6} concludes this paper.

\section{System Model}
\label{section_2}
\subsection{Neighbor discovery model}
\indent According to the assumptions of ND with directional antenna \cite{20,25}, the system model is described in detail as follows.\\
\begin{figure}[htbp]
	\centering
	\includegraphics[width=0.45\textwidth]{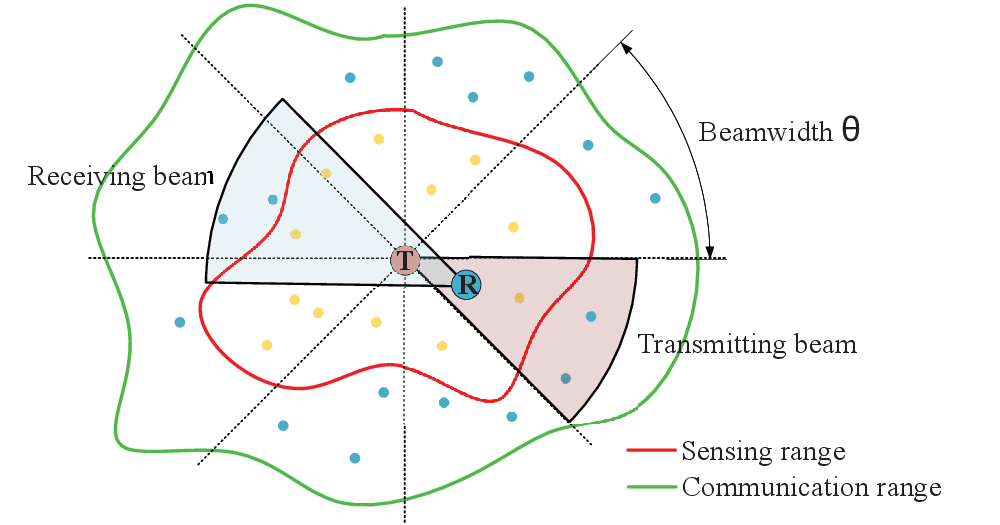}
	\caption{System model.}
	\label{fig17}
\end{figure}
\subsubsection{Unique node ID}
\indent Each node can be distinguished by a unique identifier, such as a MAC address.
\subsubsection{Handshake method}
\indent The nodes adopt the two-way handshake method. The node can discover its neighbors by receiving a hello packet or a feedback packet.
\subsubsection{Antenna model}
\indent In order to improve the capacity and extend the distance of communication, the nodes are implemented with directional antennas that operate in the Directional Transmitting and Receiving mode (DTDR). The beamwidth of the directional antenna is defined as $\theta $ $(\theta  \in (0,2\pi ])$. When $\theta  = 2\pi $, the directional antenna case can be extended to the omnidirectional antenna case. Therefore, the algorithms in this paper are also applicable to the omnidirectional antenna case. In DTDR mode, ND process requires that one node is in the transmitting state while the other is in the receiving state and ${\theta _T} = \left( {{\theta _R} + \pi } \right)\bmod 2\pi $, where ${\theta _T}$ is the transmitting angle and ${\theta _R}$ is the receiving angle, as shown in \textcolor[rgb]{0,0.4471,0.4039}{Fig. \ref{fig17}}.
\subsubsection{Time synchronization mechanism}
\indent The nodes are in the synchronous half-duplex operation mode. The time slot is taken as the unified clock, and each time slot is divided into two sub-time slots. All nodes have a fixed transmission power.
\subsubsection{Collision mechanism}
\indent A collision will occur if a node receives two or more packets at the same time.
\subsubsection{Communication range}
\indent All nodes are within a one-hop communication range of each other \cite{12}, as shown in \textcolor[rgb]{0,0.4471,0.4039}{Fig. \ref{fig17}}. Nodes are randomly distributed, and the locations of nodes in each beam follow the poisson point process.
\subsubsection{Prior mechanism}
\indent The sensing information obtained by ISAC is used as the prior information during ND. If the radar resolution is low, the node can only determine whether nodes exist or not in each beam. If the radar resolution is high, the node can obtain the number of its neighbors in each beam direction.
\subsubsection{Convergence condition}
\begin{figure}[!h]
	\centering
	\includegraphics[width=0.5\textwidth]{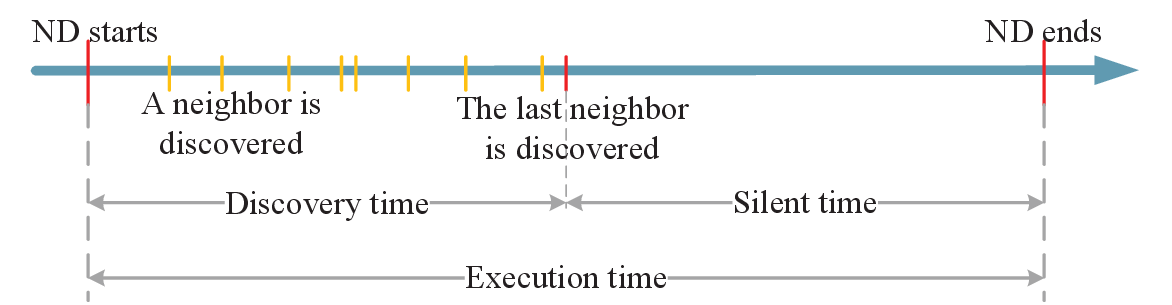}
	\caption{The process of ND.}
	\label{fig18}
\end{figure}

\indent It achieves convergence when no neighbors are discovered for a time equal to the half of the execution time of ND, as shown in \textcolor[rgb]{0,0.4471,0.4039}{Fig. \ref{fig18}}.

\subsection{Sensing model of ISAC}
The design of the sensing model faces a significant technical challenge, which lies in the formulation of the ISAC signal. To fulfill the specific requirements of the scenario, the ISAC signal can fuse the radar and communication signals, simultaneously achieving both functions. However, when employing the ISAC waveform, the radar echo signal undergoes a two-way path, which results in a reduced sensing range compared to the communication range. As a consequence, the acquired sensing information is inherently imperfect. Furthermore, imperfect sensing information can arise from factors such as node mobility and obstructions caused by obstacles.

\begin{figure}[htbp]
	\centering
	\includegraphics[width=0.45\textwidth]{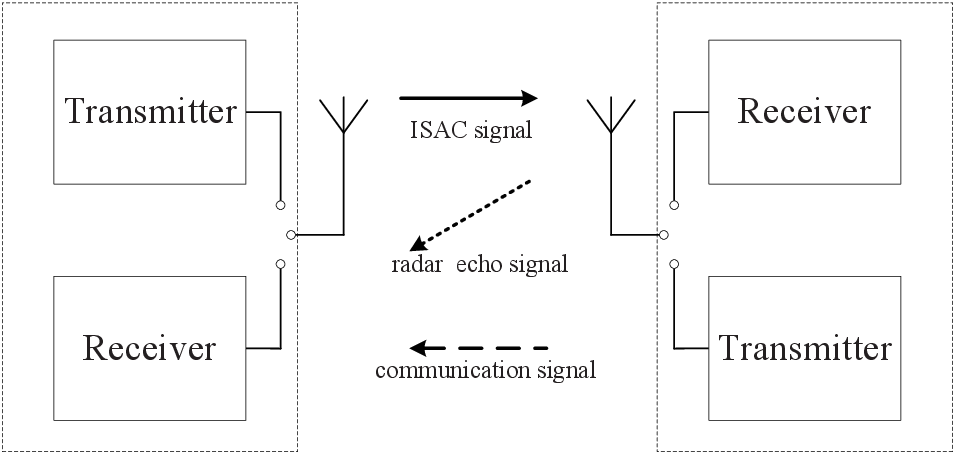}
	\caption{Sensing model of ISAC.}
	\label{fig19}
\end{figure}

The node enters the sensing mode of ISAC, randomly selects the beam direction, transmits the ISAC signal with detection capability and hello information simultaneously, and processes the radar echo signal and feedback data of the communication signal, as shown in \textcolor[rgb]{0,0.4471,0.4039}{Fig. \ref{fig19}}. It does not exit the sensing mode until each beam direction has been detected.

Each node maintains a Radar List (RL), a Neighbor List (NL) and a Communication List (CL) during the ND. Assume that there are $N$ neighbors around node $i$, randomly distributed in $k$ beams. Its RL is $\boldsymbol{\chi _i} = \{ \chi _i^1,\chi _i^2,\chi _i^3, \ldots  \ldots ,\chi _i^k\} $, where $\chi _i^m$ is the number of nodes in the $m$th beam of node $i$, which is determinded by radar signal processing. The NL is defined as $\boldsymbol{\gamma _i} = \{ \gamma _i^1,\gamma _i^2,\gamma _i^3, \ldots  \ldots \gamma _i^N\}$ $(\gamma _i^j \in \{ 0,1\} )$, where $\gamma _i^j=1$ indicates that node $i$ has discovered node $j$ and $\gamma _i^j=0$ indicates that node $j$ has not yet been discovered. In addition, the NL maintained by each node also contains the location information of its discovered neighbors. The CL of node $i$ can be generated directly from the NL. The CL is defined as $\boldsymbol{\eta _i} = \{ \eta _i^1,\eta _i^2,\eta _i^3, \ldots  \ldots ,\eta _i^k\} $, where $\eta _i^m$ is the number of discovered neighbors through communication in the $m$th beam by node $i$.

\section{Integrated Sensing and Communication Neighbor Discovery with Gossip Mechanism}
\label{section_3}

In order to achieve fast convergence of ND, the following three acceleration mechanisms are designed in this paper. 

\begin{figure}[htbp]
	\centering
	\includegraphics[width=0.45\textwidth]{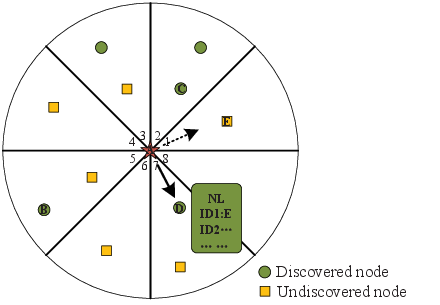}
	\caption{Acceleration mechanism of ND.}
	\label{fig2}
\end{figure}

\begin{enumerate}
	\item The non-reply mechanism: If two nodes have discovered each other before, one node will not reply even if it receives a hello package from the other node, which reduces the probability of collision \cite{27}. As shown in \textcolor[rgb]{0,0.4471,0.4039}{Fig. \ref{fig2}}, with the non-reply mechanism, node $B$ will not reply even if node $A$ transmits the hello packet to node $B$ again. The non-reply mechanism requires accurate demodulation information for the Hello packet from the transmitting node. Therefore, it's essential to demodulate the packet as correctly as possible when introducing the non-reply mechanism to the ND algorithms. 
	\item The stop mechanism: If the number of nodes in each beam is estimated using the prior information, a node will not implement ND in the completely discovered beam that covers no undiscovered neighbors \cite{27}. As shown in \textcolor[rgb]{0,0.4471,0.4039}{Fig. \ref{fig2}}, with the stop mechanism, since node $A$ has discovered all its neighbors in the beam covering node $C$, node $A$ will no longer select this beam for ND. The stop mechanism requires an estimation of the number of nodes in each beam based on prior information obtained from the radar echo signal. Therefore, it is crucial to maintain high radar sensing accuracy when introducing the stop mechanism to the ND algorithms. 
	\item The gossip mechanism: Once one node receives the feedback packet from another node, it will traverse the NL carried in the feedback packet and augment undiscovered neighbor information to its own NL to complete indirect ND. As shown in \textcolor[rgb]{0,0.4471,0.4039}{Fig. \ref{fig2}}, with the gossip mechanism, node $A$ can update its own NL by traversing the NL of node $D$, indirectly discovering node $E$.
\end{enumerate}

This paper first introduces the gossip mechanism based on the CRA algorithm. To further reduce the convergence time of the ND algorithms, we have introduced one or both of the non-reply and stop mechanisms into the gossip-based ND algorithm. As a result, this paper proposes four ISAC ND algorithms with the gossip mechanism: the non-Reply and Stop Algorithm based on Gossip (G-nRS), the Reply and non-Stop Algorithm based on Gossip (G-RnS), the non-Reply and non-Stop Algorithm based on Gossip (G-nRnS), and the Reply and Stop Algorithm based on Gossip (G-RS). The usage conditions of these three mechanisms determine the application scenarios of the four ISAC ND algorithms. The G-nRS algorithm has strict requirements for application scenarios, while the G-RnS algorithm has wider application scenarios. The G-nRnS algorithm is limited by packet demodulation accuracy, while the G-RS algorithm is constrained by radar sensing accuracy.

This section describes the algorithm design and performance analysis of these algorithms, deriving the average number of discovered nodes within a given period.

\subsection{Algorithm design and performance analysis of G-nRS}
\subsubsection{Algorithm design of G-nRS}
\begin{figure}[htbp]
	\centering
	\includegraphics[width=0.5\textwidth]{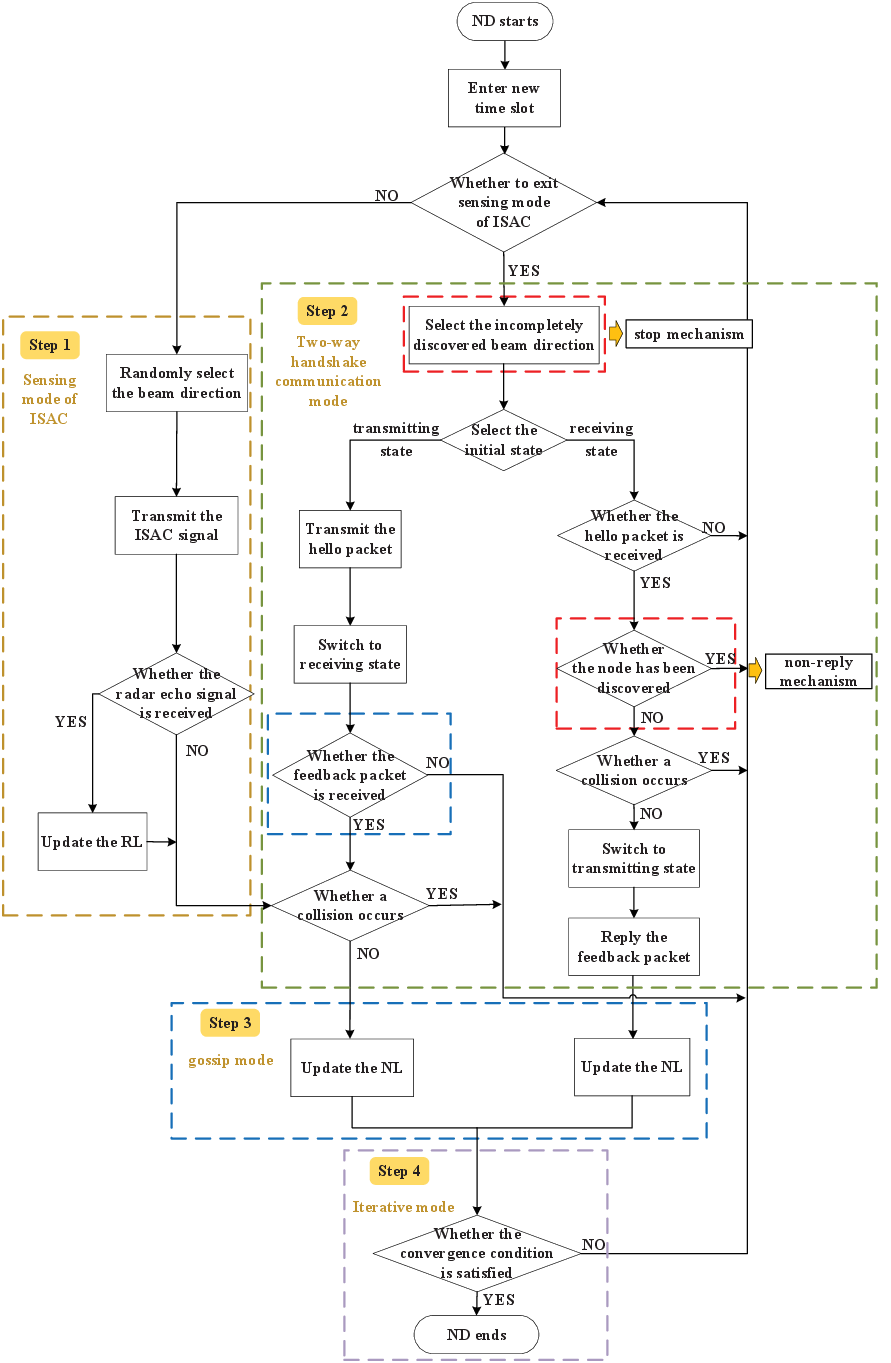}
	\caption{G-nRS algorithm flow chart.}
	\label{fig3}
\end{figure}

The G-nRS algorithm introduces the non-reply mechanism, the stop mechanism, and the gossip mechanism based on the CRA algorithm. The algorithm demonstrates outstanding convergence performance but necessitates high packet demodulation accuracy and radar sensing accuracy to achieve optimal results. The flowchart of the G-nRS algorithm is shown in \textcolor[rgb]{0,0.4471,0.4039}{Fig. \ref{fig3}}.

\begin{itemize}
	\item[$\bullet$]\textbf{\emph{Step 1}}: Each node enters the sensing mode of ISAC. In this mode, each node continuously acquires prior information until it obtains a complete RL.
	\item[$\bullet$]\textbf{\emph{Step 2}}: Each node enters the two-way handshake communication mode, scheduling the antenna to select incompletely discovered beam that covers undiscovered neighbors to transmit/receive the hello packet. To reduce the probability of collision, once two nodes have discovered each other before, they will no longer reply to the feedback packet.
	\item[$\bullet$]\textbf{\emph{Step 3}}: The nodes that successfully discover new neighbors will update their NLs with the gossip mechanism.
	\item[$\bullet$]\textbf{\emph{Step 4}}: Iterations of the above steps are carried out until the convergence condition metioned in Section \ref{section_2} is satisfied.
\end{itemize}

\subsubsection{Performance analysis of G-nRS}
\indent The average number of discovered nodes within a given period is derived as the performance metric of ND. The probability of node $i$ discovering the specific node $j$ in the $t$th time slot is
\begin{equation}
	{P_{ij}}^{\left( t \right)} = {P_{i \to j}}^{\left( t \right)} + {P_{j \to i}}^{\left( t \right)}
	\label{eq1},
\end{equation}
where ${P_{i \to j}}^{\left( t \right)}$ denotes the probability of successful ND between node $i$ in the transmitting state and node $j$ in the receiving state. It needs to satisfy the conditions that the beam directions of node $i$ and node $j$ are aligned and no interfering nodes appear in the receiving beams of node $i$ and node $j$.
\begin{equation}
	{P_{i \to j}}^{\left( t \right)} = {P_{align}}^{\left( t \right)}{P_{{T_i}}}^{\left( t \right)}{P_{{R_j}}}^{\left( t \right)}
	\label{eq2},
\end{equation}
where ${P_{align}}^{\left( t \right)}$ denotes the probability that node $i$ and node $j$ are aligned with each other in the $t$th time slot, ${P_{{T_i}}}^{\left( t \right)}$ denotes the probability that no interfering nodes appear in the receiving beam of node $i$ in the $t$th time slot and ${P_{{R_j}}}^{\left( t \right)}$ denotes the probability that no interfering nodes appear in the receiving beam of node $j$ in the $t$th time slot.\\
\indent With the stop mechanism, the completely discovered beam will not be selected again. Therefore, ${P_{align}}^{\left( t \right)}$ is
\begin{equation}
	{P_{align}}^{\left( t \right)} = {P_0}\left( {1 - {P_0}} \right)\left( {\frac{1}{{B - B_{t - 1}^i}}} \right)\left( {\frac{1}{{B - B_{t - 1}^j}}} \right)
	\label{eq3},
\end{equation}
where ${P_0}$ denotes the transmit probability of each node, $B$ denotes the number of non-empty beams that cover some neighbors and $B_{t - 1}^i$ is the number of completely discovered beams of node $i$ by $(t-1)$th time slot. \\
\indent The probability density function ${P_{N,u}}\ (u = 0,1,2, \ldots  \ldots ,N)$ of the $u$-node beam that covers $u$ neighbors is derived in \cite{27}, where $N$ denotes total number of neighbors. Thus, $B = {B_0}\left( {1 - {P_{N,0}}} \right)$, where ${B_0} = 2\pi /\theta $ is the number of beams of each node.\\
\indent With the non-reply mechanism and the stop mechanism, only part of nodes in the receiving beam of node $i$ may interfere with the reception of the feedback packet. As long as this part of nodes do not select exactly the same state and beam direction as node $j$ at this time slot, they will not become interfering nodes. Therefore, ${P_{{T_i}}}^{\left( t \right)}$ is
\begin{equation}
	{P_{{T_i}}}^{\left( t \right)} = {\left[ {1 - \left( {1 - {P_0}} \right)\frac{1}{{B - B_{t - 1}^{equal}}}} \right]^{{a_i}}}
	\label{eq4},
\end{equation} 
where ${B_{t - 1}^{equal}}$ denotes the average value of completely discovered beams of all nodes by $(t-1)$th time slot and ${a_i}$ denotes the number of nodes in the receiving beam of node $i$ that may interfere with the reception of the feedback packet.
\begin{equation}
	B_{t - 1}^{equal} = {{\sum\limits_{h = 1}^N {B_{t - 1}^h} } \mathord{\left/
			{\vphantom {{\sum\limits_{h = 1}^N {B_{t - 1}^h} } N}} \right.
			\kern-\nulldelimiterspace} N}
	\label{eq5}.
\end{equation}
\indent The nodes in the receiving beam of node $i$ may cause interference only if they have not yet discovered node $i$ and have not yet discovered all nodes in the beam covering $i$. Therefore, ${a_i}$ is
\begin{equation}
	{a_i} = \left[ {{M_i} - 1 - {U_i}\left( {t - 1} \right)} \right]{P_{emp}}\left( t \right)
	\label{eq6},
\end{equation}
where ${M_i}$ represents the number of nodes in the receiving beam of node $i$, ${U_i}\left( {t - 1} \right)$ represents the number of nodes that have discovered node $i$ in the beam covering node $j$ by $(t - 1)$th time slot and ${P_{emp}}\left( t \right)$ represents the probability that nodes in the receiving beam of node $i$ have not yet discovered all nodes in their beam covering $i$ and is given by
\begin{equation}
	{P_{emp}}\left( t \right) = 1 - \frac{{\left( \begin{array}{c}
				B - 1\\
				B_{t - 1}^{equal} - 1
			\end{array} \right)}}{{\left( \begin{array}{c}
				B\\
				B_{t - 1}^{equal}
			\end{array} \right)}} = \frac{{B - B_{t - 1}^{equal}}}{B}
	\label{eq7}.
\end{equation}
\indent The derivation process of ${P_{{R_j}}}^{\left( t \right)}$ is the same as that of ${P_{{T_i}}}^{\left( t \right)}$ and is given by
\begin{equation}
	{P_{{R_j}}}^{\left( t \right)} = {\left( {1 - {P_0}\frac{1}{{B - B_{t - 1}^{equal}}}} \right)^{{b_j}}}
	\label{eq8},
\end{equation}
where ${b_j}$ denotes the number of nodes in the receiving beam of node $j$ that may interfere with the reception of the feedback packet and is given by
\begin{equation}
	{b_j} = \left( {{M_j} - 1} \right){P_{emp}}\left( t \right)
	\label{eq9}.
\end{equation}
\indent The gossip mechanism is introduced below \cite{39}. The probability of node $i$ discovering the specific node $j$ by $t$th time slot is
\begin{equation}
	{P_{ij}}\left( t \right) = {D_{ij}}\left( t \right) + \left[ {1 - {D_{ij}}\left( t \right)} \right]{I_{ij}}\left( t \right)
	\label{eq10},
\end{equation}
where ${D_{ij}}\left( t \right)$ represents the probability that node $i$ discovers node $j$ directly by $t$th time slot and ${I_{ij}}\left( t \right)$ represents the probability that node $i$ discovers node $j$ indirectly by $t$th time slot. Then, ${D_{ij}}\left( t \right)$ and ${I_{ij}}\left( t \right)$ are formulated as
\begin{equation}
	{D_{ij}}\left( t \right) = 1 - \prod\limits_{g = 0}^t {\left( {1 - {P_{ij}}^{\left( g \right)}} \right)}
	\label{eq11},
\end{equation}
\begin{equation}
	{I_{ij}}\left( t \right) = {I_{ij}}\left( {t - 1} \right) + \left[ {1 - {I_{ij}}\left( {t - 1} \right)} \right]{A_{ij}}\left( t \right)
	\label{eq12},
\end{equation}
where ${A_{ij}}\left( t \right)$ denotes the probability that node $i$ indirectly discovers node $j$ in the $t$th time slot and is derived as
\begin{equation}
	\begin{array}{l}
		\begin{aligned}
			{A_{ij}}\left( t \right) &= \sum\limits_{m \ne i,j} {{D_{mj}}\left( {t - 1} \right)} P_{im}^{\left( t \right)}\\
			&+ \sum\limits_{m \ne i,j} {\left[ {1 - {D_{mj}}\left( {t - 1} \right)} \right]{I_{mj}}\left( {t - 1} \right)} P_{im}^{\left( t \right)}
		\end{aligned}
	\end{array}
	\label{eq13},
\end{equation}
where node $m$ is one of undiscovered common neighbors of node $i$ and node $j$.\\
\indent The boundary condition for the above iterative equation is given by
\begin{equation}
	{I_{ij}}(1) = 0
	\label{eq14}.
\end{equation}
\indent With ${P_{ij}}(t)$, ${N_i}(t)$ denoting the average number of nodes discovered by node $i$ by $t$th time slot is given by
\begin{equation}
	{N_i}\left( t \right) = E\left( {\sum\limits_{j = 1}^{N - 1} {{H_j}\left( t \right)} } \right) = {P_{ij}}\left( t \right) \times \left( {N - 1} \right)
	\label{eq15},
\end{equation}
where ${H_j}\left( t \right)$ is the event: node $j$ has been discovered by node $i$ and is given by
\begin{equation}
{H_j}\left( t \right) = \left\{ \begin{array}{l}
	\begin{array}{*{20}{c}}
		{1}&{with\ probability\ {P_{ij}}\left( t \right)}
	\end{array}\\
	\begin{array}{*{20}{c}}
		{0}&{with\ probability\ 1 - {P_{ij}}\left( t \right)}
	\end{array}
\end{array} \right.
\label{eq16}.
\end{equation}

\subsection{Algorithm design and performance analysis of G-RnS}
\subsubsection{Algorithm design of G-RnS}
\begin{figure}[htb]
	\centering
	\subfigure[The non-reply mechanism is replaced with the reply mechanism.]{
		\begin{minipage}{0.4\textwidth}
			\centering
			\label{fig5}
			\includegraphics[width=1\textwidth]{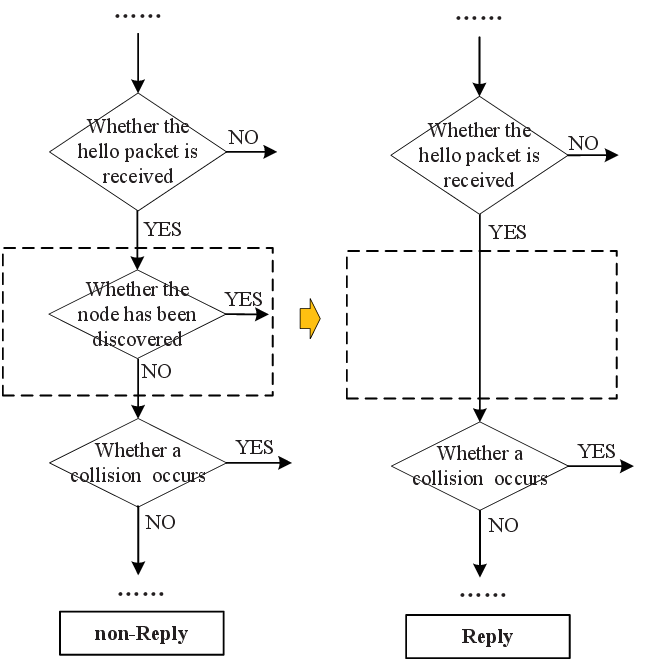}
		\end{minipage}
	}
	\vspace{0.8cm}
	\subfigure[The stop mechanism is replaced with the non-stop mechanism.]{
		\begin{minipage}{0.5\textwidth}
			\centering
			\label{fig6}
			\includegraphics[width=1\textwidth]{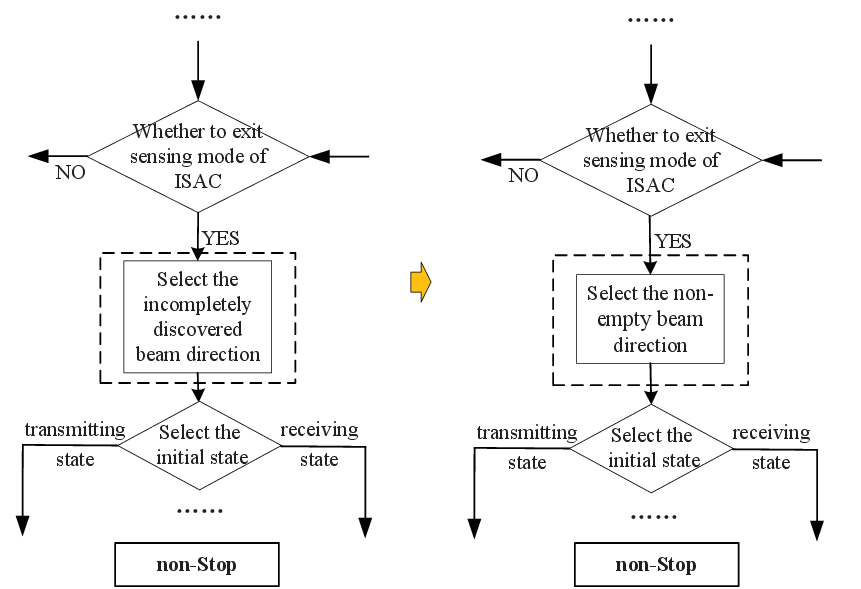}
		\end{minipage}
	}
	\caption{Modified part of G-RnS algorithm flow chart (relative to G-nRS algorithm).} \label{fig4}
\end{figure}

\indent The G-RnS algorithm introduces the gossip mechanism based on the CRA algorithm and filters out empty beams that cover no neighbors. This means that only non-empty beam directions are selected for packet transmission or reception during the ND process. The algorithm exhibits moderate convergence performance but does not impose any specific requirements on the accuracy of packet demodulation or radar sensing. Its flow chart has undergone two modifications relative to the G-nRS algorithm, as shown in \textcolor[rgb]{0,0.4471,0.4039}{Fig. \ref{fig4}}. As the simplest algorithm of the four, it requires low radar resolution because it only needs to sense the presence or absence of nodes in the beam.
\subsubsection{Performance analysis of G-RnS}
\indent The probability of a given node $i$ discovering a specific neighbor $j$ in any time slot is
\begin{equation}
	\begin{array}{l}
	\begin{aligned}
	{P_{ij}} &= {P_{i \to j}} + {P_{j \to i}}\\
	&= \frac{{\left( {1 - {P_0}} \right){P_0}}}{{{B^2}}}{\left( {1 - \frac{{1 - {P_0}}}{B}} \right)^{{M_i} - 1}}{\left( {1 - \frac{{{P_0}}}{B}} \right)^{{M_j} - 1}}\\
	&+ \frac{{\left( {1 - {P_0}} \right){P_0}}}{{{B^2}}}{\left( {1 - \frac{{1 - {P_0}}}{B}} \right)^{{M_j} - 1}}{\left( {1 - \frac{{{P_0}}}{B}} \right)^{{M_i} - 1}}
	\end{aligned}
	\end{array}
	\label{eq16}.
\end{equation}
\indent In order to make the formula simple and universal, ${M_i}$ and ${M_j}$ are substituted by average value $M$ and $M$ is given by
\begin{equation}
	M = \sum\limits_{u = 1}^N {u{P_{N,u}}'}
	\label{eq17},
\end{equation}
where ${P_{N,u}}'$ is the remaining probability density function of non-empty beams and needs to be rederived as
\begin{equation}
	{P_{N,u}}' = \frac{{{P_{N,u}}}}{{1 - {P_{N,0}}}} \quad (u = 1,2, \ldots  \ldots ,N)
	\label{eq18}.
\end{equation}
\indent Therefore, ${P_{ij}}$ is homogenized to $P$, where $P$ denotes the average probability of any node discovering a new node in any time slot and is given by
\begin{equation}
\;P = \frac{{2\left( {1 - {P_0}} \right){P_0}}}{{{B^2}}}{\left( {1 - \frac{{1 - {P_0}}}{B}} \right)^{M - 1}}{\left( {1 - \frac{{{P_0}}}{B}} \right)^{M - 1}}
	\label{eq19}.
\end{equation}
\indent The probability that any node directly discovers a new node by the $t$th time slot is
\begin{equation}
	D\left( t \right) = 1 - {\left( {1 - \ P } \right)^t}
	\label{eq20},
\end{equation}
which is a function only with time.\\
\indent With iteration, the probability that any node indirectly discovers a new node by the $t$th time slot is
\begin{equation}
	\begin{array}{l}
		\begin{aligned}
			I(t) &= \sum\limits_m { - P{{(1 - P)}^{t - 1}}} {I^2}(t - 1)\\
			&{\rm{      }} + \left\{ {\left[ {\sum\limits_m {\left( {2P{{\left( {1 - P} \right)}^{t - 1}} - P} \right)} } \right] + 1} \right\}I(t - 1)\\
			&{\rm{      }} + \sum\limits_m {P[1 - {{(1 - P)}^{t - 1}}]} 
		\end{aligned}
	\end{array}
	\label{eq21},
\end{equation}
which has been transformed into a one-dimensional quadratic function of $I(t - 1)$.\\
\indent Therefore, the average probability of any node discovering a new node by $t$th time slot is
\begin{equation}
	P(t) = D(t) + [1 - D(t)]I(t){\rm{ = }}1 - {(1 - P)^t} + {(1 - P)^t}I(t)
	\label{eq22},
\end{equation}
which is general.\\
\indent Similarly to Eq.~\eqref{eq15}, $N(t)$ denoting the average number of neighbors discovered by any node by $t$th time slot is given by
\begin{equation}
	N(t) = P(t) \times (N - 1)
	\label{eq23}.
\end{equation}
\subsection{Algorithm design and performance analysis of G-nRnS}
\subsubsection{Algorithm design of G-nRnS}
\indent The G-nRnS algorithm introduces both the non-reply mechanism and the gossip mechanism based on the CRA algorithm. The algorithm demonstrates satisfactory convergence performance. However, it requires high packet demodulation accuracy. Its flow chart has undergone a modification relative to the G-nRS algorithm, as shown in \textcolor[rgb]{0,0.4471,0.4039}{Fig. \ref{fig5}}.\\ 
\subsubsection{Performance analysis of G-nRnS}
\indent The probability of a given node $i$ discovering a specific neighbor $j$ in the $t$th time slot is
	\begin{equation}
		\begin{array}{l}
			\begin{aligned}
	{P_{ij}}^{\left( t \right)} &= {P_{i \to j}}^{\left( t \right)} + {P_{j \to i}}^{\left( t \right)}\\
	&= \frac{{\left( {1 - {P_0}} \right){P_0}}}{{{B^2}}}{\left( {1 - \frac{{1 - {P_0}}}{B}} \right)^{{M_i} - 1 - {U_i}(t - 1)}}{\left( {1 - \frac{{{P_0}}}{B}} \right)^{{M_j} - 1}}\\
	&+ \frac{{\left( {1 - {P_0}} \right){P_0}}}{{{B^2}}}{\left( {1 - \frac{{1 - {P_0}}}{B}} \right)^{{M_j} - 1 - {U_j}(t - 1)}}{\left( {1 - \frac{{{P_0}}}{B}} \right)^{{M_i} - 1}}
			\end{aligned}
		\end{array}
		\label{eq24}.
	\end{equation}
\indent In ND process, the change in ${{U_i}(t - 1)}$ and ${{U_j}(t - 1)}$ over time leads to an exponential change in ${P_{ij}}^{\left( t \right)}$ over time. However, when $B$ is large, $\left( {1 - \frac{{1 - {P_0}}}{B}} \right)$ is close to 1 and ${P_{ij}}^{\left( t \right)}$ can be approximated as a linear change. Therefore, to make the calculation process simple, ${P_{ij}}^{\left( t \right)}$ is homogenized separately in time and space \cite{27}.\\
\indent ${P_{ij}}^{\left( t \right)}$ is time-homogenized as
\begin{equation}
	\begin{array}{l}
		\begin{aligned}
			\overline {{P_{ij}}}  &= \overline {{P_{i \to j}}}  + \overline {{P_{j \to i}}} \\
			&= \frac{1}{{{T_i}}}\sum\limits_{u = 0}^{{M_i} - 1} {\int_{{T_u}}^{{T_{u + 1}}} {{P_{i \to j}}^{\left( t \right)}} } dt + \frac{1}{{{T_j}}}\sum\limits_{u = 0}^{{M_j} - 1} {\int_{{T_u}}^{{T_{u + 1}}} {{P_{j \to i}}^{\left( t \right)}} } dt
		\end{aligned}
	\end{array}
	\label{eq25},
\end{equation}
where ${T_i}$ denotes the total time to discover all nodes in the receiving beam of node $i$ and ${T_{u + 1} - {T_u}}$ denotes the time taken to discover the $u$th node in the receiving beam.\\
\indent $\overline {{P_{ij}}}$ is spatial-homogenized as
\begin{equation}
	\begin{array}{l}
		\begin{aligned}
			\overline P  = \sum\limits_{{M_i} = 1}^N {\overline {{P_{i \to j}}} }  \times {P_{N,{M_i}}} + \sum\limits_{{M_j} = 1}^N {\overline {{P_{j \to i}}} }  \times {P_{N,{M_j}}}
		\end{aligned}
	\end{array}
	\label{eq26}.
\end{equation}
\indent ${\overline P }$ is homogenized for both time and space, whose operation enhances independence and stability. Therefore, the subsequent derivation is the same as G-RnS algorithm. 

\subsection{Algorithm design and performance analysis of G-RS}
\subsubsection{Algorithm design of G-RS}
\indent The G-RS algorithm introduces both the stop mechanism and the gossip mechanism based on the CRA algorithm. The algorithm shows good convergence performance. However, it requires high radar sensing accuracy. Its flow chart has undergone a modification relative to the G-nRS algorithm, as shown in \textcolor[rgb]{0,0.4471,0.4039}{Fig. \ref{fig6}}. \\
\subsubsection{Performance analysis of G-RS}
\indent The probability of a given node $i$ discovering a specific neighbor $j$ in the $t$th time slot is\\
\begin{equation}
	\begin{array}{l}
		\begin{aligned}
	{P_{ij}}^{\left( t \right)} &= {P_{i \to j}}^{\left( t \right)} + {P_{i \to j}}^{\left( t \right)}\\
	&= {P_{align}}^{\left( t \right)}{\left( {1 - \frac{{1 - {P_{\rm{0}}}}}{{B - B_{t - 1}^{equal}}}} \right)^{{b_i}}}{\left( {1 - \frac{{{P_{\rm{0}}}}}{{B - B_{t - 1}^{equal}}}} \right)^{{b_j}}}\\
	&+ {P_{align}}^{\left( t \right)}{\left( {1 - \frac{{1 - {P_{\rm{0}}}}}{{B - B_{t - 1}^{equal}}}} \right)^{{b_j}}}{\left( {1 - \frac{{{P_{\rm{0}}}}}{{B - B_{t - 1}^{equal}}}} \right)^{{b_i}}}
		\end{aligned}
	\end{array}
	\label{eq27}.
\end{equation}
\indent To make the derivation generalizable, the following homogenization is performed.\\
\indent $B_{t - 1}^i$ and $B_{t - 1}^j$ are replaced with the average $B_{t - 1}^{equal}$, thus ${P_{align}}^{\left( t \right)}$ is homogenizated as
\begin{equation}
\overline {{P_{align}}}  = 2{P_0}(1 - {P_0}){(\frac{1}{{B - B_{t - 1}^{equal}}})^2}
	\label{eq28}.
\end{equation}
\indent ${M_i}$ and ${M_j}$ are replaced with the average $M$, thus ${b_i}$ and ${b_j}$ are homogenizated as
\begin{equation}
	{b_{equal}} = (M - 1){P_{emp}}(t)
	\label{eq29}.
\end{equation}
\indent Therefore, ${P_{ij}}^{\left( t \right)}$ is is homogenized to $\overline {{P^{\left( t \right)}}} $, where $\overline {{P^{\left( t \right)}}} $ denotes the average probability that a new node is discovered by any node in the $t$th time slot and is given by 
\begin{equation}
\overline {{P^{(t)}}}  = \overline {{P_{align}}} {\left( {1 - \frac{{1 - {P_{\rm{0}}}}}{{B - B_{t - 1}^{equal}}}} \right)^{{b_{equal}}}}{(1 - \frac{{{P_{\rm{0}}}}}{{B - B_{t - 1}^{equal}}})^{{b_{equal}}}}
	\label{eq30},
\end{equation}
which is only a function of time. Therefore, it only needs to be time-homogenized as
\begin{equation}
	\overline P  = \frac{1}{{{T}}}\sum\limits_{b = 0}^{B - 1} {\int_{{T_b}}^{{T_{b + 1}}} {\overline {{P^{\left( t \right)}}} } dt}
	\label{eq31},
\end{equation}
where ${{T}}$ denotes the total time to discover all nodes, ${T_{b + 1} - {T_b}}$ denotes the time taken to discover all nodes in the $b$th beam.\\
\indent The subsequent derivation is the same as G-RnS algorithm and the above simplified procedure makes the derivation general.
\section{Integrated Sensing and Communication Neighbor Discovery with Gossip Mechanism and Reinforcement Learning}
\label{section_4}

\subsection{Q-learning-based Neighbor Discovery Model}
\indent In specific scenarios, various factors, including disparities between sensing and communication ranges, node mobility, and obstruction caused by obstacles, collectively contribute to the imperfection of sensing information. This imperfect sensing information hampers the accurate acquisition of the neighboring node distribution. Under the restriction of the convergence condition, the presence of imperfect sensing information may result in incomplete ND and the convergence delay. Incomplete ND represents the existence of undiscovered nodes within the communication range, even after the ND process has reached convergence. This situation gives rise to potential issues such as incomplete routing information and degradation in network performance. To mitigate the impact caused by imperfect sensing information, this paper substitutes the stop mechanism that required high accuracy of sensing information with the Q-learning mechanism, proposing the non-Reply and non-Stop Algorithm based on Gossip and Q-learning (GQ-nRnS).

\indent Each node acts as an agent with learning ability. The introduction of the reinforcement learning mechanism enables agents to supplement the prior information through interaction with the environment, reducing the number of discarded neighbor nodes due to imperfect sensing information and accelerating convergence, which makes the ND process flexible, efficient, and scalable.

\begin{figure}[htbp]
	\centering
	\includegraphics[width=0.4\textwidth]{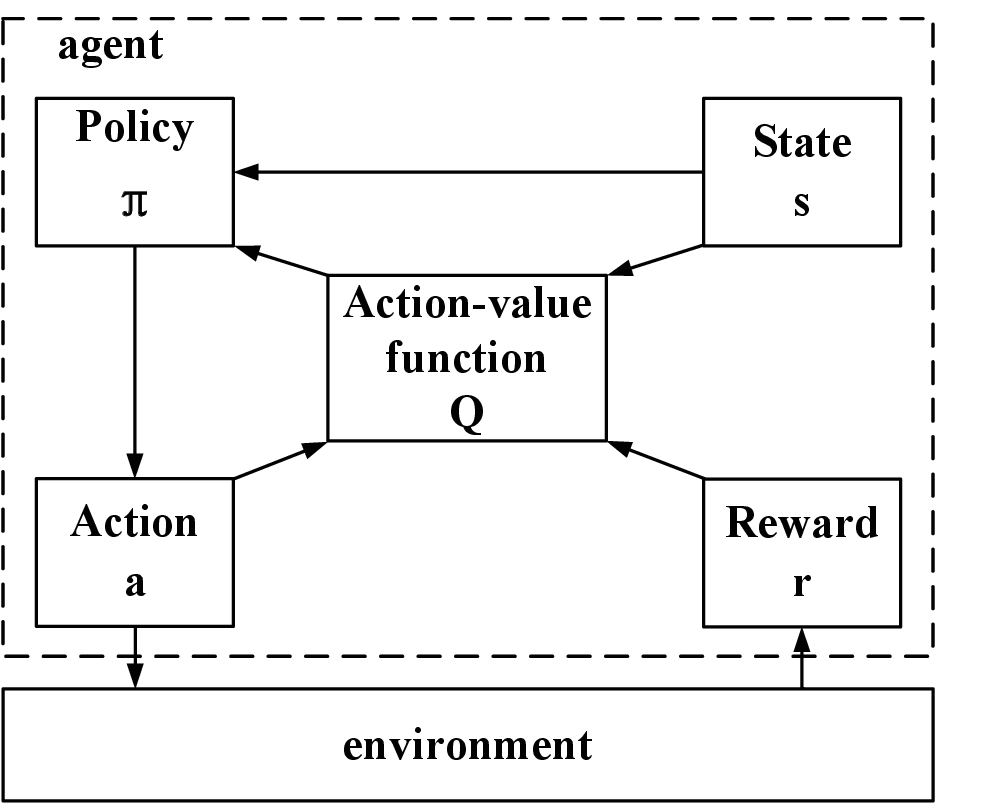}
	\caption{Interaction between the agent and the environment.}
	\label{fig7}
\end{figure}

\indent The interaction between the agent and the environment is shown in \textcolor[rgb]{0,0.4471,0.4039}{Fig. \ref{fig7}}. Through the interaction with the environment, the agent obtains information such as neighbor location, number of neighbors, etc.. Comparing environment information with the prior information, the agent can acquire the reward $r$ in the current state $s$ ,which is used for updating the value function $Q$. Subsequently, the agent will select corresponding action $a$ under the instruction of strategy $\pi $ to conduct the ND. The steps above are cycled until the convergence condition metioned in Section \ref{section_2} is satisfied.\\
\subsubsection{State}
\indent The selection of state variables is required to be memoryless. Since the priori information in GQ-nRnS algorithm will influence antenna beam direction selection, the antenna transceiver state is determined as the state $s$. The state space is defined as $S = \{ S(0),S(1)\} $, where $S(1)$ denotes the transmitting state and $S(0)$ denotes the receiving state.
\subsubsection{Action}
\indent The antenna beam direction is determined as the action $a$. The antenna beam width is set to $\theta $, so that there are $B = 2\pi /\theta $ different beam directions. The action space is defined as $A = \{ {A_1},{A_2},{A_3}, \ldots  \ldots ,{A_B}\} $, where ${A_k}$ denotes $k$th beam direction.
\subsubsection{Reward}
\indent In each time slot, the agent compares RL and CL to obtain the reward $r$. When RL is greater than CL, the reward is positive, which indicates that there are still undiscovered nodes in this beam. When RL is less than or equal to CL, it will be discussed in two cases, which is depended on the average value of nodes in each beam. 

\begin{figure}[!h]
	\centering
	\includegraphics[width=0.45\textwidth]{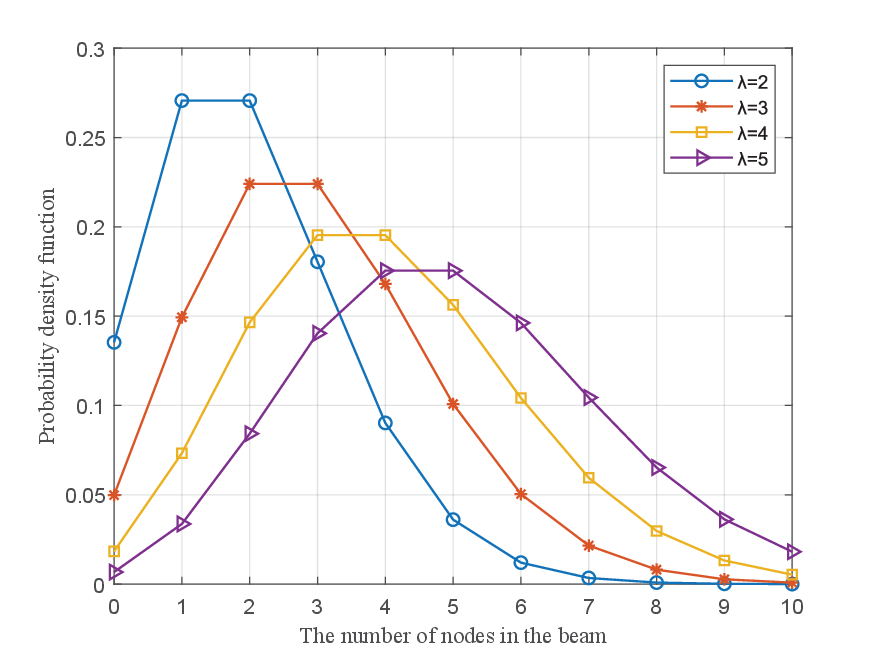}
	\caption{Probability density function of nodes in each beam.}
	\label{fig16}
\end{figure}

Assume that the locations of nodes in each beam follows a poisson point process with density $\lambda $, as shown in \textcolor[rgb]{0,0.4471,0.4039}{Fig. \ref{fig16}}, where $\lambda $ denotes the average value of nodes in each beam.

When CL is less than or equal to the Extreme Point (EP), the reward is positive. Since there is a high probability that there are some potential nodes that are not sensed and undiscovered in this beam. When CL is greater than EP, the reward is negative. Because the probability that there are potential nodes in this beam decreases with the increase of the discrepancy between CL and EP. To achieve the best tradeoff between convergence accuracy and convergence speed, the reward $r$ is defined as \cite{34}
\begin{equation}
r\left( {s,a} \right) = \left\{ {\begin{array}{*{20}{l}}
		2&{{\rm{RL > CL}}}\\
		1&{{\rm{RL}} \le {\rm{CL}},{\rm{CL}} \le {\rm{EP}}}\\
		{ - 1}&{{\rm{RL}} \le {\rm{CL}},{\rm{CL}} > {\rm{EP}}}
\end{array}} \right.
	\label{eq34}.
\end{equation}
\indent The dynamic reward puts agents into an adaptive mode to solve the convergence delay caused by imperfect prior information.
\subsubsection{Action-value function}
\indent The goal of the intelligent node is to find the best strategy to maximize gains \cite{47}. The action-value function $Q$ represents the total payoff and is given by \cite{48} 
\begin{equation}
	Q(s,a) \leftarrow Q(s,a) + \alpha [R + \gamma \mathop {\max }\limits_{a'} Q(s',a') - Q(s,a)]
	\label{eq35},
\end{equation}
where $\alpha  \in [0,1]$ is the learning rate and $\gamma  \in [0,1]$ is the discount factor.

\subsubsection{Policy finding algorithm}
\indent GQ-nRnS algorithm adopts $\varepsilon  - greedy$ policy finding algorithm to achieve the trade-off between the ``exploitation'' and ``exploration'' \cite{49}. The agent exploits the optimal action with probability $1 - \varepsilon $ to approach the optimal strategy gradually and explores non-optimal action with probability $\varepsilon $ to avoid falling into sub-optimal policies. In addition, the exploration coefficient $\varepsilon $ gradually decreases over time, so that the initial value $\varepsilon $ is particularly important. The action $a$ is obtained as
\begin{equation}
a = \left\{ {\begin{array}{*{20}{l}}
		{\mathop {\max }\limits_{a'} Q(s,a')}&{{\rm{with}}\;{\rm{probability}}\;1 - \varepsilon }\\
		{{\rm{random}}}&{{\rm{with}}\;{\rm{probability}}\;\varepsilon }
\end{array}} \right.
	\label{eq36}.
\end{equation}
\subsection{Algorithm flow}
The operational flow of the ISAC ND with gossip mechanism and reinforcement learning is as \textcolor[rgb]{0,0.4471,0.4039}{Algorithm \ref{alg1}}.

\begin{algorithm}[!h]
	\caption{GQ-nRnS algorithm flow.}\label{alg:alg1}
	\begin{algorithmic}[1]
		\STATE \textbf{Initialize global variables.}
		\STATE Initialize the locations of $N$ nodes;
		\STATE Initialize RL and CL of each node as an $N \times B$ zero matrix; 
		\STATE Initialize NL of each node as an $N \times N$ zero matrix;
		\STATE Initialize $Q$ of each node as an $2 \times B$ zero matrix;
		\STATE \textbf{Enter the sensing mode of ISAC to obtain complete RLs.}
		\REPEAT
		\STATE \textbf{Enter the two-way handshake communication mode.}
		\STATE Generate the random number $greedy \in [0,1]$.
		\FOR{int $i=1$ to $N$}
		\STATE Initialize the local variable reward $r$.
		\IF{$sum(Q(:)) =  = 0$}
		\STATE The agent randomly selects the action $a_i \in A$.
		\ELSE
		\IF{$greedy < \varepsilon$}
		\STATE The agent explores an action $a_i \in A$ with a non-maximum $Q$.
		\ELSE
		\STATE The agent exploits the action $a_i \in A$ with the maximum $Q$.
		\ENDIF
		\ENDIF
		\STATE Continue with the subsequent ND.
		\STATE According to Eq.~\eqref{eq34}, obtain reward $r\left( {s_i,a_i} \right) $.
		\STATE According to Eq.~\eqref{eq35}, update function $Q(s_i,a_i)$. 
		\ENDFOR
		\STATE \textbf{Go to the next time slot and update the state ${s^{\left( t \right)}} \leftarrow {s^{\left( {t + 1} \right)}}$.}
		\UNTIL{the convergence condition metioned in Section \ref{section_2} is satisfied}.
	\end{algorithmic}
	\label{alg1}
\end{algorithm}
\section{Numerical Results}
\label{section_5}

\subsection{G-RnS, G-nRnS, G-RS, G-nRS algorithms}
\indent It is assumed that there are 50 nodes randomly deployed in a 2 km $ \times $ 2 km square area. Each node is equipped with a manipulable directional antenna whose beamwidth is fixed at 14.4\degree. The communication range and radar sensing range are both set to $2\sqrt 2$ km. The effectiveness of the four algorithms, G-RnS, G-nRnS, G-RS, and G-nRS, is quantified by two widely used performance metrics as follows. 

\begin{itemize} 
	\item ND ratio over time: The ratio of discovered neighbors by the certain time slot to the total number of nodes in the network. 
	\item Convergence time: The number of time slots required to satisfy the convergence condition metioned in Section \ref{section_2}.
\end{itemize}

\indent In the following simulations, G-nRS algorithm is selected as an example to prove the effectiveness of the theoretical results.
\subsubsection{G-nRS algorithm}

\begin{figure}[!h]
	\centering
	\includegraphics[width=0.45\textwidth]{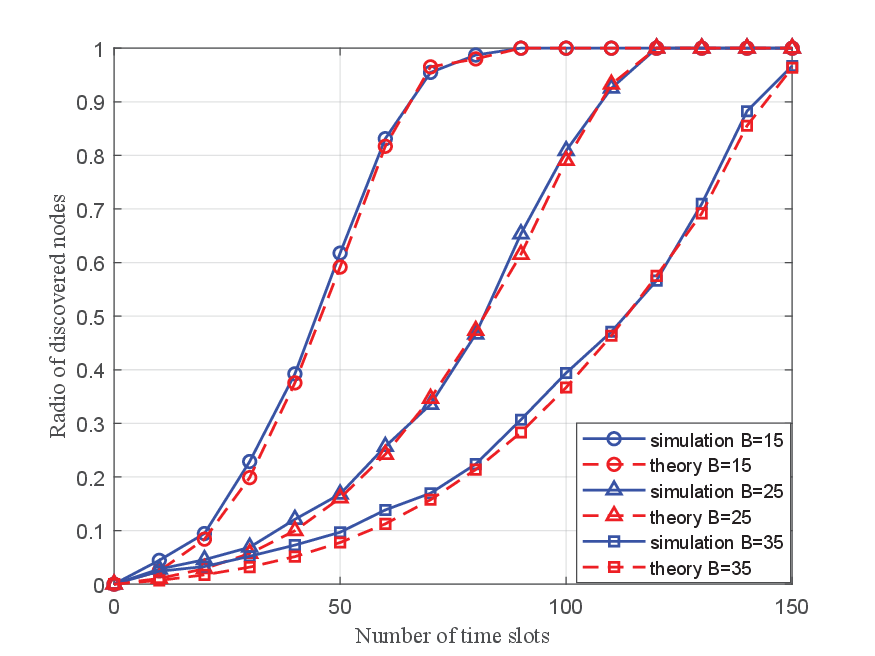}
	\caption{Theoretical results and simulation results with G-nRS algorithm under different number of beams.}
	\label{fig8}
\end{figure}

Section \ref{section_3} derives the relationship between the number of discovered nodes and the number of required time slots of G-nRS algorithm. As shown in \textcolor[rgb]{0,0.4471,0.4039}{Fig. \ref{fig8}}, it is revealed that the theoretical results and simulation results fit well, which can verify the correctness of the theoretical results of G-nRS algorithm.

\subsubsection{G-nRS, G-RnS, G-nRnS, and G-RS algorithms}

\begin{figure}[!h]
	\centering
	\includegraphics[width=0.45\textwidth]{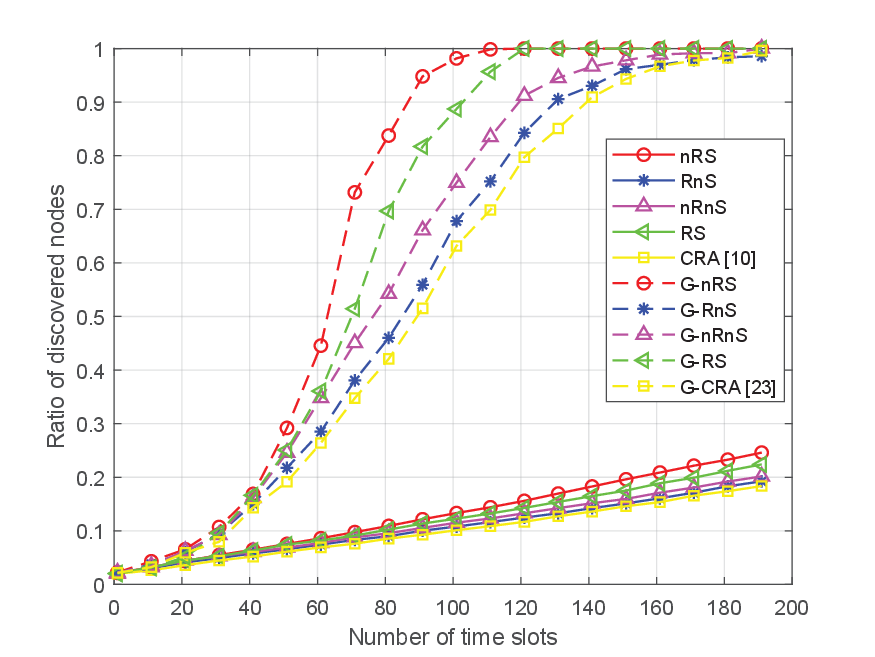}
	\caption{ND radio over time with ten algorithms.}
	\label{fig9}
\end{figure}

\indent \textcolor[rgb]{0,0.4471,0.4039}{Fig. \ref{fig9}} simulates the ND ratio over time with ten algorithms, which are non-Reply and Stop Algorithm (nRS), Reply and non-Stop Algorithm (RnS), non-Reply and non-Stop Algorithm (nRnS), Reply and Stop Algorithm (RS), CRA \cite{12}, G-nRS, G-RnS, G-nRnS, G-RS, and Completely Random Algorithm based on Gossip (G-CRA) \cite{39}. Compared to the RnS algorithm, nRnS algorithm, RS algorithm, and nRS algorithm, the G-RnS algorithm, G-nRnS algorithm, G-RS algorithm, and G-nRS algorithm only introduce the gossip mechanism based on the above algorithms. As a result, their convergence time has been reduced by 96.85$\%$, 96.65$\%$, 87.07$\%$, and 87.21$\%$, respectively. It is revealed that the gossip mechanism has the best acceleration effect, and the advantage becomes more and more distinct over time. Besides, G-nRS algorithm significantly outperforms the remaining nine algorithms, whose ND efficiency is improved by about 98.23$\%$ compared with CRA algorithm and about 51.88$\%$ compared with G-CRA algorithm. It is shown that the interplay between the priori mechanism and the gossip mechanism further facilitates fast convergence of ND algorithms.

\begin{figure}[!h]
	\centering
	\includegraphics[width=0.45\textwidth]{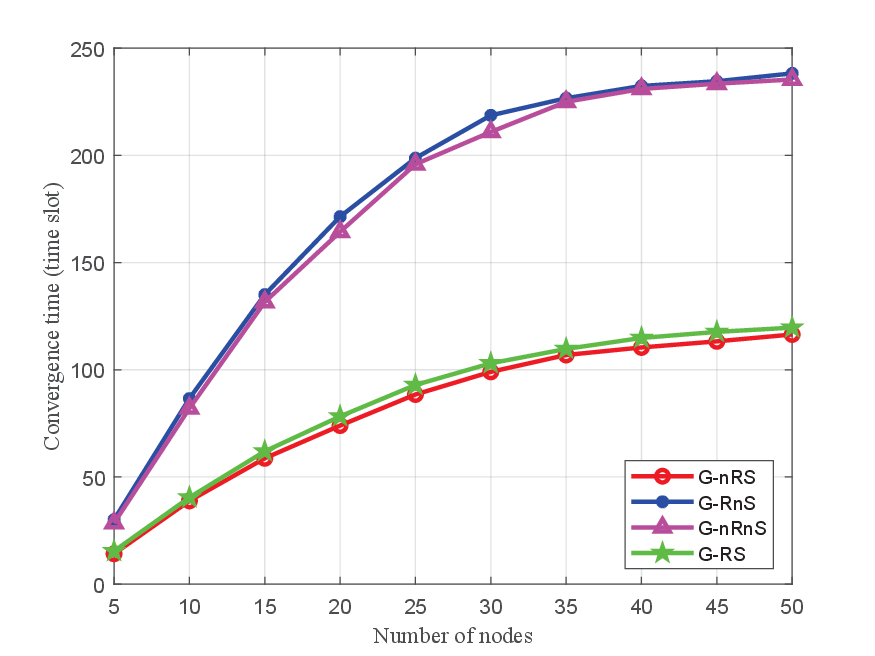}
	\caption{Convergence time with G-RnS, G-nRnS, G-RS, G-nRS.}
	\label{fig10}
\end{figure}

\indent \textcolor[rgb]{0,0.4471,0.4039}{Fig. \ref{fig10}} shows the average convergence time of G-nRS, G-RnS, G-nRnS, and G-RS when the number of nodes varies from 5 to 50. When the number of nodes is 50, the convergence time of G-nRS is decreased by 51.12$\%$, 50.53$\%$, and 2.67$\%$ compared with G-RnS, G-nRnS, and G-RS. It is revealed that, in gossip-based ND algorithms, the utilization of the stop mechanism leads to significantly faster convergence compared to the non-reply mechanism.

\subsection{GQ-nRnS algorithm}
\indent It is assumed that the number of nodes varies from 5 to 50, randomly deployed in a 2 km $\times$ 2 km square area. Each node is equipped with a manipulable directional antenna whose beamwidth is fixed at 36\degree. The communication range is set to $2\sqrt 2$ km. The radar sensing range is initially set to $2\sqrt 2$ km and modified according to the $R - C{\rm{ }}$ $ratio$, which is used to simulate the cases with imperfect sensing information.\\
\indent Since ``exploration'' process in GQ-nRnS algorithm lead to high uncertainty on the number of discovered nodes in a time slot, the ND ratio over time cannot accurately reveal variation characteristics. Therefore, the simulation of GQ-nRnS algorithm focuses on the average convergence time.\\

\subsubsection{GQ-nRnS algorithm under different parameters}

\begin{figure}[htbp]
	\centering
	\includegraphics[width=0.45\textwidth]{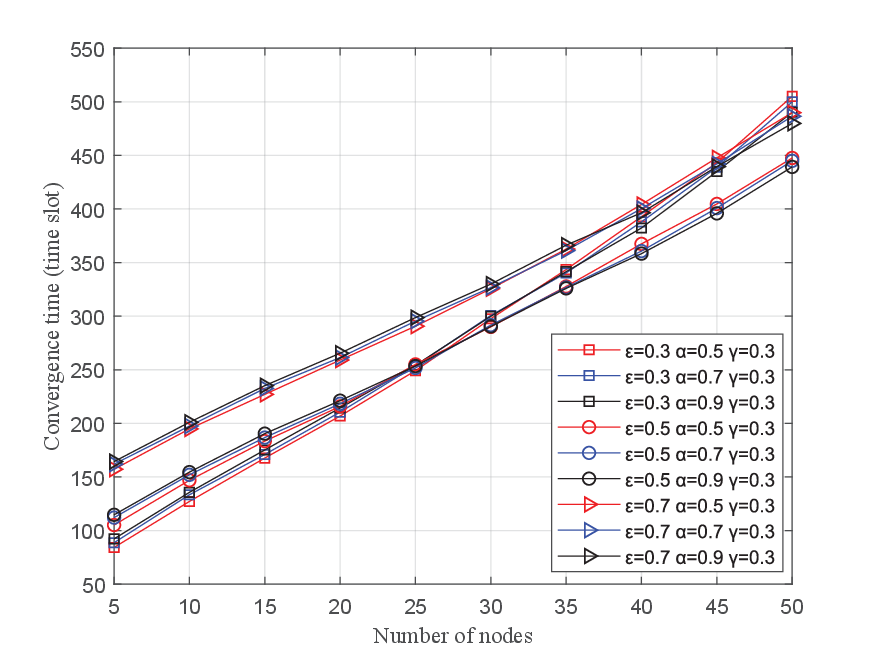}
	\caption{Convergence time with GQ-nRnS algorithm under different parameters.}
	\label{fig11}
\end{figure}

\indent \textcolor[rgb]{0,0.4471,0.4039}{Fig. \ref{fig11}} investigates the effect of different parameters on the performance of GQ-nRnS algorithm. Since the exploration coefficient $\varepsilon $ is a decreasing function of time, the $\varepsilon $ in the simulation is the initial value. To eliminate the impact of radar sensing distance on the performance of ND algorithms, the $R - C{\rm{ }}$ $ratio$ is set to 1.

\indent The role of $\gamma $ is to strike a balance between short-term and long-term gains. Because of the independence between the processes of discovering each potential neighbor node, the long-term goal of discovering all potential neighbors can be divided into the short-term goal of discovering potential neighbors within each time slot. The gains obtained from the short-term and long-term goals are consistent, indicating that the weighting factor $\gamma $ has little or even negligible impact on the convergence performance of the algorithm. Numerous simulation results also confirm the reasonableness of the above analysis. {\textcolor[rgb]{0,0.4471,0.4039}{Fig. \ref{fig11}}} sets $\gamma $ as a constant and proceeds to analyze the effect of $\varepsilon $ and $\alpha $ on the performance of the algorithm. As shown in {\textcolor[rgb]{0,0.4471,0.4039}{Fig. \ref{fig11}}}, the selection of exploration coefficients $\varepsilon $ directly affects the algorithm performance. When the number of nodes ranges from 5 to 25, GQ-nRnS scheme with $\varepsilon  = 0.3$ performs best. When the number of nodes ranges from 25 to 50, GQ-nRnS scheme with $\varepsilon  = 0.5$ performs best. It is revealed that as the number of nodes increases, more ``exploration" is needed to avoid falling into a sub-optimal strategy. Furthermore, the learning rate $\alpha $ also has a weak impact on algorithm performance. As the number of nodes increases, the algorithm performance is improving with increasing $\alpha $. Since as the density of nodes increases, the experience gained in each time slot is important for the selection of the best action.\\

\subsubsection{GQ-nRnS algorithm under different $R - C{\rm{ }}$ $ratio$}

\begin{figure}[htbp]
	\centering
	\includegraphics[width=0.45\textwidth]{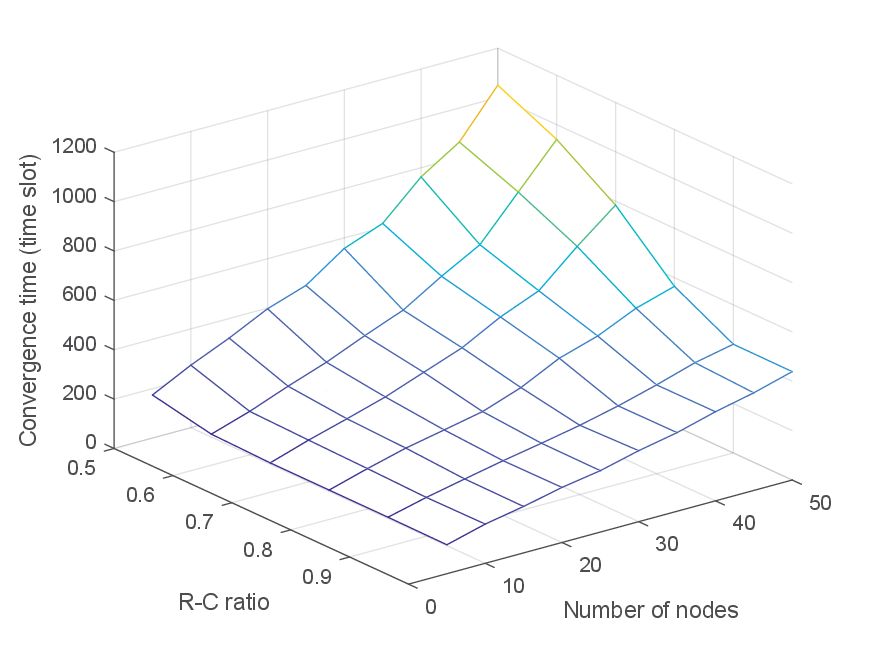}
	\caption{Convergence time with GQ-nRnS algorithm under different $R - C{\rm{ }}$ $ratio$ and number of nodes.}
	\label{fig14}
\end{figure}

\textcolor[rgb]{0,0.4471,0.4039}{Fig. \ref{fig14}} explores the impact of the $R - C{\rm{ }}$ $ratio$ on the average convergence time under different number of nodes in GQ-nRnS algorithm, where parameters in simulation are $\varepsilon  = 0.5$, $\alpha  = 0.9$, $\gamma  = 0.5$.\\
\indent As shown in \textcolor[rgb]{0,0.4471,0.4039}{Fig. \ref{fig14}}, the number of time slots increases with the decrease of $R - C{\rm{ }}$ $ratio$. The reason is that the accuracy of the prior information is enhancing with the increase of $R - C{\rm{ }}$ $ratio$, which accelerates the convergence of ND efficiently. When $R - C{\rm{ }}$ $radio = 0.5,0.8,1$ respectively, the average convergence time for 50 nodes is 83.4$\%$, 78.9$\%$, 73.3$\%$ higher than that for 5 nodes. The increase in the accuracy of the priori information slows down the growth of convergence time caused by the increased node density. The reason is that when the number of nodes increases, the probability of collision between nodes is increasing, and the improvement in the accuracy of the prior information makes the non-reply mechanism effective in reducing the collision probability. Finally, when $R - C{\rm{ }}$ $radio \ge 0.9$, the surface is relatively steady, and the performance tends to be stable.

\subsubsection{Q-ND, Q-nR, GQ-ND, and GQ-nRnS algorithms}

\begin{figure}[htbp]
	\centering
	\includegraphics[width=0.45\textwidth]{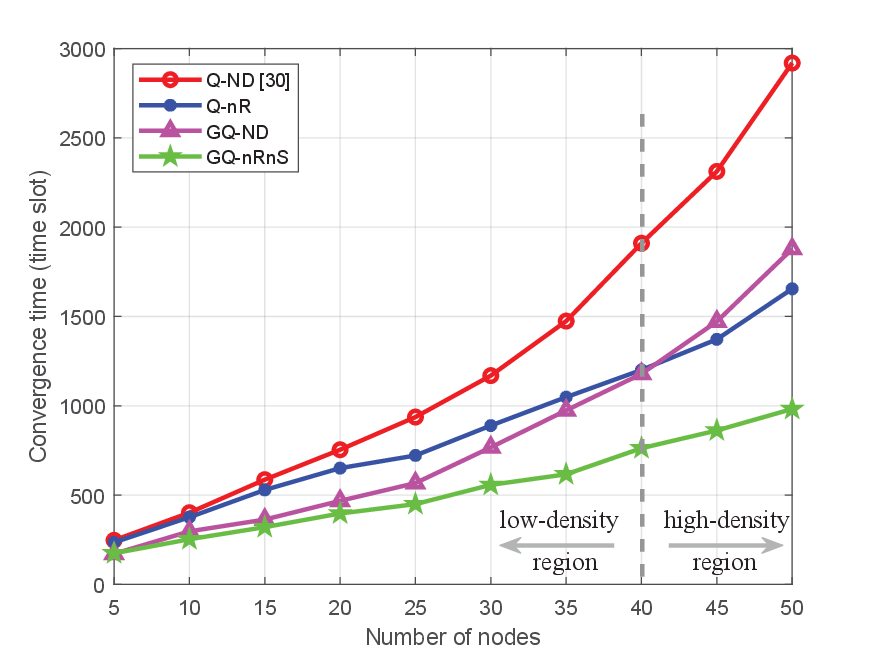}
	\caption{Convergence time with Q-ND, Q-nR, GQ-ND, and GQ-nRnS.}
	\label{fig15}
\end{figure}

\indent \textcolor[rgb]{0,0.4471,0.4039}{Fig. \ref{fig15}} shows the relationship between the average convergence time and the number of nodes for the four algorithms, which are the Q-learning-based ND Algorithm (Q-ND) \cite{34}, the Q-learning-based ND Algorithm with non-Reply mechanism (Q-nR), the Q-learning-based ND Algorithm with Gossip mechanism (GQ-ND), and the GQ-nRnS algorithm. The Q-nR algorithm enhances the Q-ND algorithm by incorporating the non-reply mechanism. Similarly, the GQ-ND algorithm improves upon the Q-ND algorithm by introducing the gossip mechanism. Finally, the GQ-nRnS algorithm combines both the non-reply mechanism and the gossip mechanism, building upon the Q-ND algorithm. The $R - C{\rm{ }}$ $radio$ is set to 0.5 and the parameters are set to $\varepsilon  = 0.5$, $\gamma  = 0.3$, $\alpha  = 0.5$.\\
\indent As shown in \textcolor[rgb]{0,0.4471,0.4039}{Fig. \ref{fig15}}, it is revealed that both the prior information and the gossip mechanism can accelerate the convergence of ND algorithms based on Q-learning. The gossip mechanism performs well in low-density region while the non-Replay mechanism performs well in high-density region. Therefore, the interplay between the non-reply mechanism and the gossip mechanism can fully exploit the advantages of them in accelerating the convergence of ND algorithms. The GQ-nRnS algorithm outperforms the other three algorithms distinctly for both low density and high density regions. When the number of nodes is 50, the average convergence time of GQ-nRnS algorithm is reduced by about 66.4$\%$ compared with Q-ND algorithm. GQ-nRnS algorithm still maintains the high performance in the case of imperfect sensing information.

\section{Conclusion}
\label{section_6}
\indent This paper proposes G-nRS algorithm, G-RnS algorithm, G-nRnS algorithm, and G-RS algorithm and derives the average number of discovered nodes within a given period as the critical metric to evaluate the performance of ND algorithms. The simulation results verify the correctness of theoretical derivation. Besides, we can conduct the conclusion that the interplay between the prior mechanisms and the gossip mechanism not only reduces the information redundancy in the network, but also further reduces the convergence time of ND algorithms. In addition, in order to solve the problem of imperfect sensing information, this paper proposes GQ-nRnS algorithm. Under the constraints of the convergence condition, GQ-nRnS algorithm not only ensures the completeness of ND, but also still maintains the high convergence efficiency of ND. The convergence time of GQ-nRnS algorithm is reduced by 66.4$\%$ compared with Q-ND algorithm. 

\indent For the future work, we will further explore and employ ISAC techniques to develop efficient and adaptive algorithms for topology construction and maintenance, aiming to address the challenges posed by high node mobility and limited energy resources in MANETs.

\end{spacing}
\end{document}